\documentclass[sigconf]{acmart}

\AtBeginDocument{%
  }

\copyrightyear{2026}
\acmYear{2026}
\setcopyright{cc}
\setcctype{by}
\acmConference[Internetware 2026]{the 17th International Conference on Internetware}{July 18--20, 2026}{Gold Coast, QLD, Australia}
\acmBooktitle{the 17th International Conference on Internetware (Internetware 2026), July 18--20, 2026, Gold Coast, QLD, Australia}
\acmDOI{10.1145/3834680.3834727}
\acmISBN{979-8-4007-2891-4/2026/07}

\usepackage{graphicx}
\usepackage{booktabs}
\usepackage{caption}
\begin{document}

\title{When Should Dependency Updates Invoke Repair Agents? A Lightweight Routing Study}


\author{Liheng Fan}
\email{fanliheng0807@gmail.com}
\affiliation{%
  \institution{China Academy of Telecommunication Technology}
  \city{Beijing}
  \country{China}
}

\author{Jialun Yin}
\email{yinjialun@tsari.tsinghua.edu.cn}
\affiliation{%
  \institution{School of Vehicle and Mobility, Tsinghua University}
  \city{Beijing}
  \country{China}
}

\author{Yuzhi Chen}
\email{y.z.chen.fhy@gmail.com}
\affiliation{%
  \institution{Intelligent Transportation System Research Center, Southeast University}
  \city{Nanjing}
  \country{China}
}




\begin{abstract}
Dependency-update pull requests are frequent and mostly routine, but a small subset requires non-trivial compatibility repair. Recent repository-level coding agents make such repair increasingly plausible, yet invoking them on every dependency update wastes model calls, CI time, repository context, and review attention. We frame this as a pre-agent routing problem: deciding which dependency-update pull requests should be escalated before downstream diagnosis or repair attempts. We introduce DepFixRouter, a lightweight router that ranks dependency updates by historical compatibility-repair likelihood using creation-time textual and metadata signals. On 497 labeled GitHub dependency-update candidates, only 72 require substantive repair. A creation-time-safe LinearSVC using only PR titles and bot/dependency flags reaches 0.488 repair F1 and captures 51.4\% of repairs within the top 20\% routed pull requests, improving calls per captured repair from 6.90 under route-all or random policies to 2.68. Retrospective full-history signals improve top-20\% recall to 65.3\%, revealing substantial hindsight leakage in pull-request histories rather than deployment-time routing utility. In a 60-case diagnosis-agent pilot, router-gated diagnosis reduces actual LLM calls by 66.7\% and tokens by 66.1\%, suggesting budget-aware escalation while measuring diagnosis rather than patch generation. DepFixRouter can serve as a lightweight escalation layer between routine dependency-update automation and expensive repository-level agents, enabling budget-aware maintenance without relying on retrospective repair evidence for deployment-time routing.

\end{abstract}


\maketitle

\section{Introduction}

Dependency-update automation has become part of routine software maintenance. Bots such as Dependabot and Renovate continuously submit pull requests that update package versions, regenerate lockfiles, and modify workflow dependencies~\cite{he2023dependabot,githubDependabotDocs,renovateDocs}. Many of these pull requests are intentionally small, yet this apparent simplicity hides an important maintenance problem. Prior work has shown that dependency updates can introduce breaking changes, compilation failures, and test failures that require non-trivial developer intervention~\cite{kula2018dependencies,bump2024,reyes2026byam}. Developers may need to adjust build scripts, pin or roll back transitive dependencies, rewrite tests, update type annotations, or adapt source code to changed APIs. These repairs are not always obvious from the initial version bump and often appear as follow-up commits such as ``fix tests,'' ``fix lint,'' ``pin eslint,'' or ``adapt to new API.'' The resulting research question is whether a dependency-update workflow can identify likely repair cases before invoking expensive repository-level repair machinery.

LLM-based coding agents provide a plausible mechanism for handling such cases, especially as recent software-engineering benchmarks and agents frame repository-level issue resolution as an interactive code-editing task~\cite{swebench2024,sweagent2024}. As software-engineering agents become more capable, the scarce resource shifts from patch generation alone to deciding when expensive repository-level reasoning should be invoked. This places dependency-update repair at the intersection of software-engineering agents, cost-aware AI systems, and Internetware-style adaptive software maintenance. However, using an agent on every dependency update is a poor default because it consumes model calls, CI time, repository context, and developer review attention, while existing dependency-update workflows and repair-agent evaluations still do not resolve this pre-agent routing problem. Because repair-requiring updates are sparse in our repair-enriched sample and likely sparse in routine update streams, an all-agent strategy may spend substantial effort on low-yield cases.

Existing work addresses adjacent parts of this problem but leaves the pre-agent routing question underexplored. Dependency-update studies characterize bot behavior and update failures, breaking-update benchmarks such as BUMP support reproducible repair evaluation, and LLM-based repair work such as Byam studies how models fix already observed dependency breakages. Repository-level agents and agentless pipelines further show that models can navigate codebases, edit files, and run tools~\cite{sweagent2024,openhands2025,agentless2025}, while cost-aware LLM routing shows that expensive models should be invoked selectively~\cite{chen2023frugalgpt,ong2025routellm}. What remains unclear is the preceding decision point: before running a repository-level diagnosis or repair agent, can a dependency-update workflow identify which pull requests are likely to require compatibility work? This pre-repair distinction is the gap addressed by DepFixRouter.
\begin{figure*}[t]
  \centering
  \includegraphics[width=0.85\textwidth]{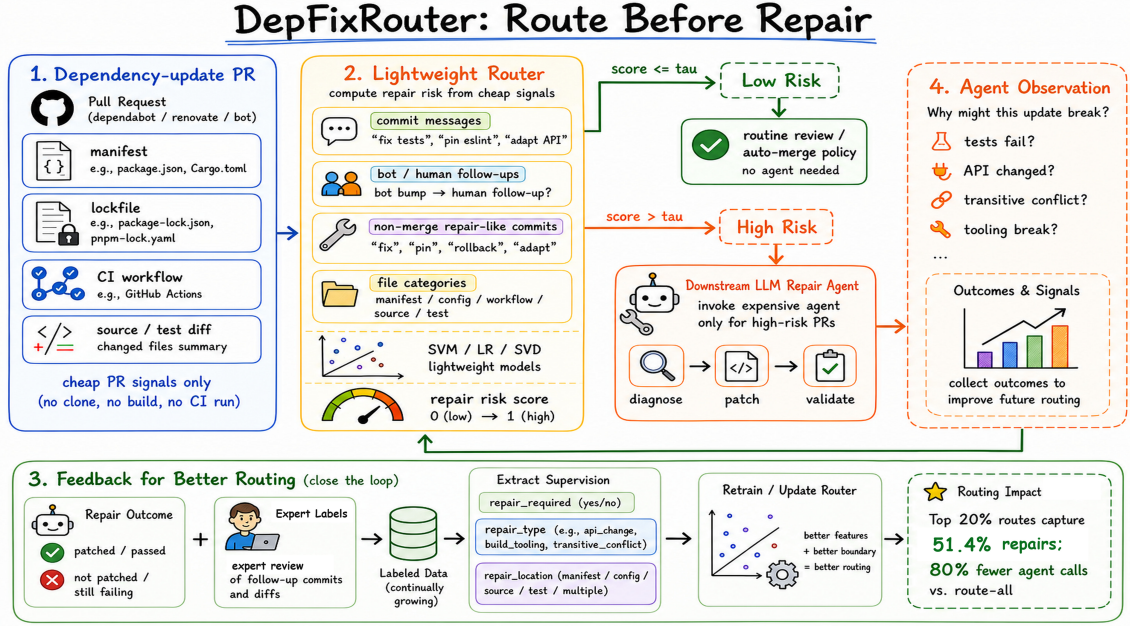}
  \caption{The proposed DepFixRouter framework. The figure illustrates the feature-availability-aware routing setup and distinguishes deployment-time creation-safe routing from richer initial-PR T0 and retrospective full-history analyses. Any retrospective utility numbers shown are diagnostic and should not be interpreted as live deployment performance.}
  \label{fig:overview}
\end{figure*}

We introduce DepFixRouter as a router-first framework for this pre-agent escalation problem. Figure~\ref{fig:overview} summarizes the framework design: inexpensive pull-request signals feed a lightweight router, the router selects a budgeted subset for downstream diagnosis or repair, and historical outcomes support later routing-policy refinement. This organization directly addresses the pre-agent routing gap by making escalation a tunable decision before expensive repository-level reasoning is invoked. We therefore evaluate DepFixRouter with matched creation-time-safe, pre-human-follow-up T0, and retrospective full-history comparisons, so that deployment-time routing evidence is separated from retrospective repair mining. Throughout the paper, ``repair risk'' denotes historical compatibility-repair likelihood, not observed LLM-agent repairability.

This paper makes three preliminary contributions. First, we formulate dependency-update repair routing as a pre-agent escalation problem for cost-aware software maintenance, shifting the focus from repair-label classification to deciding when expensive repository-level reasoning should be invoked. Second, we propose a feature-availability-aware evaluation protocol that separates creation-time-safe routing from retrospective full-history repair mining, avoiding inflated deployment claims from post-hoc pull-request evidence. Third, we provide an initial expertly labeled dataset and lightweight baselines showing that inexpensive creation-time signals can enrich repair-requiring dependency updates, while post-hoc PR traces substantially inflate apparent routing utility.

\section{Dependency-Induced Repair Triage}

\subsection{Repair Definition}

Dependency-update pull requests are noisy. A pull request may contain a bot-generated dependency bump, merge commits from the base branch, generated lockfile changes, unrelated feature work, or broad CI maintenance. Treating every non-bot commit as repair would greatly overestimate dependency-induced repair work.

In this paper, repair refers to dependency-update-induced compatibility adaptation, not necessarily bug repair in the traditional automated program repair sense. We label a pull request as requiring repair only when there is evidence that the dependency update caused a substantive follow-up change. Examples include source changes caused by a changed dependency API, test updates caused by changed behavior, configuration changes needed for newer tooling, and transitive dependency pins or rollbacks needed to restore compatibility. We do not count merge-only commits, lockfile regeneration without repair evidence, generated rebuilds, unrelated refactorings, generic CI hardening, or dependency-bot configuration changes as positive repairs.

This boundary matters because DepFixRouter is not a detector of extra commits. Non-bot follow-up events may be weak retrospective signals, but they are not sufficient labeling evidence by themselves. In this study, DepFixRouter estimates a historical repair-likelihood proxy: whether the dependency update created compatibility work that may indicate a need for downstream diagnosis or patching.

Our labels use historical human repair activity as a proxy for future repair need. A positive label means that a dependency-update pull request eventually required substantive compatibility work, not that an agent would necessarily reproduce the fix. We therefore study repair-likelihood triage rather than guaranteed agent-success prediction.

\subsection{Why Triage Comes Before Repair}

An LLM repair agent may be useful when a dependency update breaks tests, introduces type errors, or requires API migration. But when the update only changes manifests or lockfiles, invoking an agent adds cost without much expected benefit. The envisioned workflow therefore has two stages: \textbf{routing}, which decides whether a pull request is likely to require repair; and \textbf{repair}, which diagnoses the failure, edits files, and validates the patch if routed.

This makes DepFixRouter a ranked risk scorer rather than only a binary classifier. A project with limited CI resources may route only the top 10\% highest-risk updates, while a security-sensitive project may route the top 30\% or 50\%. Low-risk routing means skipping the repair agent, not automatic merging; normal review and CI policies still apply.

\section{DepFixRouter}

DepFixRouter is designed as a lightweight pre-agent router that uses inexpensive pull-request signals under scarce labels and supports adjustable routing thresholds for trading off agent-call volume against repair recall. It uses textual, metadata, and file-category signals from the pull-request record. The textual view includes the title and commit messages. Static metadata captures bot authorship and dependency-related flags, such as whether the PR appears bot-authored and whether its title or metadata indicate a dependency update.

We explicitly separate three feature-availability regimes: \textbf{minimal creation-time-safe}, \textbf{richer initial-PR T0}, and \textbf{post-hoc full history}. Minimal creation-time-safe models use only the title plus bot/dependency flags. Richer initial-PR T0 models additionally fetch initial bot commit messages and initial commit counts before any human follow-up. Full-history models add human follow-up indicators, later commit messages, mixed file counts, and other post-hoc evidence. The current router is deliberately lightweight: we explore linear SVM and logistic-regression models over TF-IDF text features, metadata features, and dense SVD projections of the text. The central claim is not that a particular classifier is optimal, but that inexpensive pull-request signals can support selective routing.

\section{Preliminary Evidence}

\subsection{Dataset and Evaluation}

We collected dependency-update repair candidates from GitHub using dependency-bot terms combined with repair-like terms such as fix, test, build, pin, rollback, and API. Candidate pull requests were manually labeled under the strict repair definition above. Labels were assigned by a single annotator with a second consistency pass over ambiguous cases. We retained dependency-update candidates and excluded unrelated maintenance PRs during expert inspection. The collection strategy intentionally targets a repair-enriched pool rather than a natural dependency-update stream. The dataset contains 500 candidates; 497 have resolved binary labels and 3 ambiguous cases are retained in the dataset count but excluded from all binary-label evaluations. Only 72 of the 497 resolved examples are positive, a 14.5\% rate even under this enriched sampling strategy.

We evaluate models with 5-fold repository-grouped cross-validation using out-of-fold risk scores so each pull request is ranked by a model that was not trained on that repository. We report macro-F1 and repair F1 for binary classification, but use top-$k$ repair recall and precision@k as the primary routing metrics. AUPRC denotes area under the precision-recall curve. For the DeepSeek baseline, we use \texttt{deepseek-chat} with temperature 0 and JSON output; the prompt includes the label definition, PR title, bot/dependency flags, changed-file categories, and full commit messages. This baseline is also retrospective because it sees full-history fields. We include DeepSeek only as an expensive retrospective comparator, not as a lightweight deployment router; its row is not a same-budget comparison with the fixed top-20\% policies.

\begin{center}
\small
\setlength{\tabcolsep}{3pt}
\captionof{table}{Creation-time-safe, richer initial-PR T0, and retrospective full-history comparison.}
\label{tab:t0}
\resizebox{\columnwidth}{!}{
\begin{tabular}{lrrrr}
\toprule
Model & Macro-F1 & Repair F1 & AUPRC & Top-20\% Rec. \\
\midrule
Creation-time: title only + LinearSVC + flags & 0.679 & 0.488 & 0.334 & 0.514 \\
Initial-PR T0: title + init commits + flags/count & 0.671 & 0.475 & 0.331 & 0.472 \\
History: title + post-hoc metadata + LinearSVC & 0.702 & 0.516 & 0.435 & 0.611 \\
History: title + all commits + LinearSVC + metadata & 0.724 & 0.551 & 0.459 & 0.653 \\
History: title+body+commits + LinearSVC + metadata & 0.720 & 0.541 & 0.504 & 0.625 \\
\bottomrule
\end{tabular}
}
\end{center}

Table~\ref{tab:t0} is the core methodological result. The minimal creation-time-safe baseline reaches 0.488 repair F1 and 0.514 top-20\% recall. Adding initial bot commit messages in the richer initial-PR T0 variant yields similar repair F1 (0.475) but lower top-20\% recall (0.472), suggesting that initial commit text is not a stable ranking improvement in this small dataset. Retrospective full-history signals remain stronger, reaching 0.551 repair F1 and 0.653 top-20\% recall in the matched title+commits comparison.  

\subsection{Routing Utility}

For repair-likelihood triage, ranking utility is more important than standalone classification accuracy because the router is intended to control when expensive downstream agents are invoked. Table~\ref{tab:routing_utility} compares fixed-budget top-20\% routing policies across feature-availability settings. All fixed-budget policies use global top-$k$ rankings over the 497 scored pull requests, so the reported calls directly correspond to the number of dependency-update pull requests that would be escalated for downstream diagnosis or repair.

The creation-time-safe top-20\% policy routes 99 of 497 pull requests and captures 37 of 72 repairs. This improves precision from the 14.5\% base rate to 37.4\%, and reduces calls per captured repair from 6.90 under route-all or random routing to 2.68. Thus, even before observing post-hoc pull-request activity, inexpensive creation-time signals provide a concrete way to reduce low-yield downstream agent invocations. The richer initial-PR T0 policy captures 34 repairs at the same call budget, while the matched full-history policy captures 47 repairs. This gap shows that retrospective pull-request histories contain substantial hindsight signal. We therefore interpret full-history results as diagnostic evidence for repair mining, not as live deployment savings.

\begin{center}
\small
\captionof{table}{Top-20\% routing utility. Random is the expected value under the base positive rate.}
\label{tab:routing_utility}
\begin{tabular*}{\columnwidth}{@{\extracolsep{\fill}}lrrrr@{}}
\toprule
Policy & Calls & Repairs & Recall & Calls/Repair \\
\midrule
Route all & 497 & 72 & 1.000 & 6.90 \\
Random 20\% & 99 & 14.4 & 0.200 & 6.90 \\
Creation-safe & 99 & 37 & 0.514 & 2.68 \\
Initial-PR T0 & 99 & 34 & 0.472 & 2.91 \\
Full history & 99 & 47 & 0.653 & 2.11 \\
\bottomrule
\end{tabular*}
\end{center}

\subsection{Measured Diagnosis-Agent Pilot}

The routing simulation estimates how many pull requests would be escalated, but it does not by itself measure actual agent cost. We therefore ran a small diagnosis-agent pilot on 60 dependency-flagged pull requests: 20 sampled from the creation-time top-20\% pool, 20 from a random pool, and 20 from the bottom-20\% pool. This pilot measures diagnosis escalation cost, not repository editing or successful patch generation. A fixed DeepSeek diagnosis prompt with temperature 0 decided whether each pull request likely required compatibility diagnosis or patching; no repository checkout, editing, or CI execution is measured.

\begin{center}
\small
\captionof{table}{Measured diagnosis-agent pilot. The endpoint is diagnosis escalation, not patch generation.}
\label{tab:agent_pilot}
\begin{tabular*}{\columnwidth}{@{\extracolsep{\fill}}lrrrr@{}}
\toprule
Policy & Calls & Tokens & TP & Calls/TP \\
\midrule
All-agent & 60 & 22,827 & 5 & 12.0 \\
Router top-20 & 20 & 7,737 & 4 & 5.0 \\
Random-20 & 20 & 7,573 & 1 & 20.0 \\
Bottom-20 & 20 & 7,517 & 0 & -- \\
\bottomrule
\end{tabular*}
\end{center}

Table~\ref{tab:agent_pilot} shows that all-agent diagnosis used 60 calls and 22,827 tokens, producing 5 true-positive escalations among 12 gold repair cases. Router-gated diagnosis on the top-20\% pool used 20 calls and 7,737 tokens, reducing calls by 66.7\% and tokens by 66.1\%. It retained 4 of the 5 true-positive escalations produced by the all-agent diagnosis policy, but covered only 4 of the 12 gold repairs in the 60-case sample. The random-20 pool retained only one true-positive escalation, and the bottom-20 pool retained none. Thus DepFixRouter improves measured diagnosis-call efficiency in this sample, although the pilot remains exploratory and does not validate end-to-end automated repair.

\section{Threats to Validity}

Temporal feature availability is the main internal threat. Some signals, such as human follow-up commits, later repair-like messages, and later file changes, are only observable after developers have already reacted to the dependency update. We therefore separate creation-time-safe routing, richer initial-PR T0 analysis, and retrospective full-history analysis. The retrospective results are useful for understanding hindsight evidence in pull-request histories, but they should not be interpreted as live routing performance.

The dataset is manually labeled but modest in size, especially for the positive class. The 14.5\% positive rate is measured inside a repair-enriched candidate pool and should not be interpreted as the global repair rate of all dependency updates. Labels rely on historical human repair activity as a proxy for future repair need, not as evidence that a current repair agent would reproduce the fix. The single-annotator protocol, repair-like search queries, and exploratory model search may introduce subjectivity, sampling bias, and model-selection bias. The diagnosis-agent pilot reports real calls, tokens, and latency, but it measures escalation only; repository checkout, patch generation, CI execution, and human review cost remain future work.

\section{Conclusion}

Dependency-update repair is sparse, costly to identify, and only partially visible at pull-request creation time. This paper addresses the pre-agent escalation problem in agentic dependency maintenance: before invoking expensive repository-level coding agents, a lightweight triage layer should estimate which dependency-update pull requests are likely to require compatibility diagnosis or patching.

We introduced DepFixRouter as an initial step toward this routing-first workflow. On 497 labeled dependency-update repair candidates, creation-time-safe signals already provided non-trivial enrichment: routing the top 20\% highest-risk pull requests captured 51.4\% of repair-requiring cases, reducing calls per captured repair from 6.90 under route-all or random routing to 2.68. Retrospective full-history signals achieved higher recall, but this improvement also revealed substantial hindsight evidence in pull-request histories. In a measured 60-case diagnosis-agent pilot, router-gated diagnosis reduced actual LLM calls by 66.7\% and tokens by 66.1\%, retaining 4 of the 5 true-positive escalations produced by all-agent diagnosis while covering 4 of the 12 gold repairs in the sample. This pilot measures diagnosis escalation cost, not repository editing or successful patch generation. The central methodological lesson is therefore to evaluate deployment-time routing separately from retrospective repair mining.

DepFixRouter is not a replacement for automated program repair or LLM-based dependency repair; it is a front-end decision layer for deciding when expensive repair machinery should be invoked. Future work should extend creation-time signals, evaluate calibrated budget-recall policies across ecosystems, and measure end-to-end utility in terms of agent success, token cost, CI time, and human review effort. More broadly, repair-likelihood routing shifts the question from only how software agents fix dependency updates to when they should be activated.

\bibliographystyle{ACM-Reference-Format}
\bibliography{references}

\end{document}